\documentclass[amsfonts,12pt]{article}%
\usepackage{amssymb,amsmath,latexsym}
\usepackage{graphicx}
\usepackage{amsmath}
\usepackage{amsfonts}
\usepackage{amssymb}%
\providecommand{\U}[1]{\protect\rule{.1in}{.1in}}
\def \ed{\end{document}}
\numberwithin{equation}{section}
\def \n1{\newpage}
\def \L1{\frak L}

\def \U{{\mathcal U}}

\def \bqn{\begin{equation}}
\def \9{\end{equation}}
\def \3{\begin{eqnarray*}}
\def \4{\end{eqnarray*}}
\def \1{\begin{eqnarray}}
\def \2{\end{eqnarray}}

\def \big1{\bigcap}

\newcounter{corollary}

\newcounter{proposition}
\newcounter{definition}

\def \brem{\begin{remark}}
\def \erem{\end{remark}}
\def \bth{\begin{theorem}}
\def \eth{\end{theorem}}
\def \bpr{\begin{proposition}}
\def \epr{\end{proposition}}
\def \bprf{\begin{proof}}
\def \eprf{\end{proof}}
\def \bex{\begin{example}}
\def \eex{\end{example}}
\def \bprf{\begin{proof}}
\def \eprf{\end{proof}}
\def \blem{\begin{lemma}}
\def \elem{\end{lemma}}
\newcounter{theorem}

\def \big{\bigcap}

\def \bl{\begin{lemma}}
\def \bcor{\begin{corollary}}
\def \ecor{\end{corollary}}
\def \el{\end{lemma}}
\def \beq*{\begin{eqnarray*}}
\def \eeq*{\end{eqnarray*}}
\def \6{\vspace*{7mm}}

\def \s1{\sqrt}

\def \bt{\begin{tabular}}
\def \et{\end{tabular}}

\def \hs1{\hspace*{3mm}}
\def \q2{\hspace*{9mm}}

\def \un1{\underline}

\def \vs1{\vspace*{4mm}}

\def \ba{\begin{array}}
\def \ea{\end{array}}

\newcommand{\ec}{\end{center}}
\newcommand{\bc}{\begin{center}}
\newcommand{\be}{\begin{equation}}

\newcommand{\ee}{\end{equation}}
\newcommand{\bn}{\begin{enumerate}}
\newcommand{\en}{\end{enumerate}}
\newcommand{\bi}{\begin{itemize}}
\newcommand{\ei}{\end{itemize}}
\newtheorem{theorem}{Theorem}

\newtheorem{corollary}{Corollary}

\newtheorem{example}{Example}

\newtheorem{lemma}{Lemma}

\newtheorem{proposition}{Proposition}
\newtheorem{remark}{Remark}

\newenvironment{proof}[1][Proof]{\textbf{#1.} }{\
\rule{0.5em}{0.5em}}
\begin{document}

\title{Polynomial solutions of certain differential equations arising in
physics}
\author{H. Azad, A. Laradji and M. T. Mustafa}
\date{}
\maketitle \vspace{-.5in}
\begin{center}
Department of Mathematics \& Statistics, King Fahd University of Petroleum \& Minerals,
Dhahran 31261, Saudi Arabia\\
\texttt{hassanaz@kfupm.edu.sa}, \texttt{alaradji@kfupm.edu.sa} and \texttt{tmustafa@kfupm.edu.sa}\\
\end{center}

\baselineskip=18pt
\begin{abstract}
Linear differential equations of arbitrary order with polynomial coefficients
are considered. Specifically, necessary and sufficient conditions for the
existence of polynomial solutions of a given degree are obtained for these
equations. An algorithm to determine these conditions and to construct the
polynomial solutions is given. The effectiveness of this algorithmic approach
is illustrated by applying it to several differential equations that arise in
mathematical physics.
\end{abstract}


\renewcommand{\theequation}{\thesection.\arabic{equation}}

\section{Introduction}

\noindent Differential equations of the form $\sum_{k=0}^{N}a_{k}y^{(k)}=0$
where $a_{k}$ is a polynomial of degree$\ \leq k$ $(1\leq k\leq N)$ have been
studied by many authors, notably Bochner \cite{bochner} and Brenke
\cite{brenke} for $N=2,$ and Krall \cite{krall} and Littlejohn
\cite{littlejohn} (see also \cite{krall littlejohn}) for orthogonal polynomial
solutions. However, the case when the polynomials $a_{k}$ have arbitrary
degree has not been investigated as extensively. In their recent paper
\cite{ciftci2010}, Ciftci et al. considered certain types of such equations
that arise in mathematical physics. They specifically gave conditions for the
existence of polynomial solutions using, in particular, the asymptotic
iteration method (AIM) they introduced in their earlier work \cite{c2003}.

\noindent In this paper we consider linear differential equations of arbitrary
order with polynomial coefficients of arbitrary degree. Our approach is based
on linear algebra and provides not only a necessary and sufficient condition
for the existence of polynomial solutions of such equations, but also an
algorithmic procedure for the verification of this condition as well as for
constructing these solutions. This is discussed in detail in Section 2. In
Section 3, we include a Maple program that can be implemented to determine the
conditions that guarantee the existence of polynomial solutions of any degree
and to find them, depending of course on the available computational power. We hope that the computer program given here will be useful for researchers in solving linear differential equations with polynomial coefficients. To
illustrate the efficiency of our algorithmic procedure, the Maple code is
implemented in several examples, namely the one-dimensional Schr\"{o}dinger
equation, planar Coulomb diamagnetic problem, Bohr Hamiltonian with Davidson
potential, and radial Schr\"{o}dinger equation with shifted potential. We
point out that in \cite{ciftci2010} the authors stated that finding polynomial
solutions is a problem that needs to be investigated and suggested AIM for that.

\section{Polynomial Solutions}

\noindent Throughout, $\mathbb{P}$ is the space of all real polynomials and
$\mathbb{P}_{n}$ is the subspace of polynomials with degree at most $n$. Let
$L:\mathbb{P}\rightarrow\mathbb{P}$ be the linear operator given
$Ly=\displaystyle\sum_{k=0}^{N}p_{k}(x)D^{k}y,$ where $D$ is the usual
differential operator and $p_{k}(x)=\sum_{h\geq0}p_{kh}x^{h}$ is a polynomial
of degree $d_{k}$ ( with the convention that the zero polynomial has degree
$-\infty$ and that $D^{0}y=y).$ Our objective is to find a necessary and
sufficient condition for the equation $Ly=0$ to have non-trivial polynomial
solutions. Although this can be achieved, for each specific case, by comparing
coefficients (see for example the determinantal necessary condition in the
recent interesting paper \cite{ciftci2010} on Heun's equations), or by using
the Asymptotic Iteration Method in the case of second-order equations
\cite{c2003}, we feel that a systematic approach that works for differential
equations of all orders and that can easily be implemented in a computer
algebra system is more desirable.\newline

\noindent Assume first that for some $i~(0\leq i\leq N),d_{i}>i.$ Let
$\displaystyle m=\max_{0\leq i\leq N}\left(  {d_{i}-i}\right)  $ and put
$y=D^{m}z.$ In this way, the equation $Ly=0$ is equivalent to $Hz=0$ where $H$
is the linear operator $\displaystyle\sum_{k=1}^{m+N}a_{k}(x)D^{K}$, and
$Ly=0$ has a polynomial solution of degree $n\geq0$ if and only if $Hz=0$ has
a polynomial solution of degree $n+m$. Clearly, for each nonnegative integer
$n,\mathbb{P}_{n}$ is $H$-invariant, and $H$ has thus the advantage over $L$
of being directly amenable to an eigenvalue analysis as demonstrated below. We
note here that the easier case when $d_{i}\leq i$ for all $i$ can be discussed
almost verbatim, with obvious modifications.\newline

\noindent Let $a_{k}$ $(k\geq1)$ be the sequence of polynomials defined by
$a_{k}=0$ if $k<m$ and $a_{k}=p_{k-m}$ if $k\geq m$. Put $a_{k}(x)=\sum
_{h\geq0}a_{kh}x^{h},$ where $a_{kh}=0$ if $k<h$. Since, for each nonnegative
integer $n,$ $H(x^{n})$ is a scalar multiple of $x^{n}$ plus lower order
terms, we see that the matrix representation of $H$, with respect to the
standard basis $B_{n}=\{1,x,...,x^{n}\}$ of $\mathbb{P}_{n}$ is upper
triangular and its eigenvalues are the coefficients of $x^{n}$ in $H(x^{n})$.
More specifically, the $(n+1)\times(n+1)$ matrix $A_{n}$ of $H$ operating on
$\mathbb{P}_{n}$ has $(i,j)$-th entry $\displaystyle\sum_{k\geq1}%
a_{k,\,k+i-j}(j-k)_{k},$ i.e.%

\[
\displaystyle A_{n}=\left[  \sum_{k\geq1} a_{k,\, k+i-j}(j-k)_{k}\right]
_{1\leq i, \, j\leq n+1}%
\]

\noindent where $(j-k)_{k}=(j-1)(j-2)\cdots(j-k),$ and where each row and
column has at most $(N+m+1)$ nonzero entries. Clearly, the first $m$ columns
of $A_{n}$ are zero and $A_{n+1}$ is obtained by $A_{n}$ by adding one row and
one column at the end. As diagonal entries of $A_{n}$, all the eigenvalues of
the operator $H$ are real and are given by $\displaystyle\lambda_{n}%
=n!\sum_{k=1}^{n}\frac{a_{kk}}{(n-k)!}$ for $n\geq1$ (note that $\lambda
_{0}=\lambda_{1}=\cdots=\lambda_{m-1}=0)$. Each eigenvalue $\lambda_{n}$ has
an eigenpolynomial $y_{n}(x)=y_{n0}+y_{n1}x+\cdots+y_{nn}x^{n}$ of degree at
most $n$ and whose vector representation $(y_{n0},...,y_{nn})^{T}$ in the
standard basis $B_{n}$ can be directly computed from the homogeneous upper
triangular system $(A_{n}-\lambda_{n}I)(y_{n0},...,y_{nn})^{T}=0.$ Our problem
is to find necessary and sufficient conditions for which the operator $H$ has
an eigenpolynomial of degree $n+m$ corresponding to $\lambda_{n+m}=0,$ that is
necessary and sufficient conditions for the homogeneous system $A_{n+m}%
(y_{n+m,\,0},...,y_{n+m,n+m})^{T}=0$ to have a solution $(y_{n+m,\,0}%
,...,y_{n+m,n+m})^{T}$ with $y_{n+m,n+m}=1.$ This will follow from \newline

\noindent\textbf{Lemma 1. }Let $A$ be an $m\times n$ matrix. Then the
homogeneous system $AX=0$ has a solution $X=(x_{1},x_{2},...,x_{n})^{T}$ with
$x_{k}\neq0$ for some $k$ if and only if $rank(A)=rank(A_{k})$ where $(A_{k})$
is the matrix obtained from $A$ by deleting the $k^{th}$\textit{
column.}\newline

\noindent\textbf{Proof. }Put $A=[c_{ij}]_{1\leq i,j\leq n}$ and let $c_{k}$ be
the $k^{th}$ column of $A$. Clearly, $A$ and the augmented matrix
$[A_{k}\vdots c_{k}]$ have the same rank. Hence,
\begin{eqnarray*}
& rank(A) = rank(A_{k})\Leftrightarrow rank[A_{k}\vdots c_{k}]=rank(A_{k}
)\Leftrightarrow\text{ the system}A_{k}X=-c_{k}\text{ is consistent}\\
& \Leftrightarrow\text{there exists a solution }X=(x_{1},\cdots,x_{k-1}
,1,x_{k+1},\cdots,x_{n})^{T}\text{ to the system }AX=0. \ \ \Box
\end{eqnarray*}

\noindent Since $A_{n+m-1}$ is obtained from $A_{n+m}$ by deleting the last
column, the above lemma immediately yields that the differential equation
$Ly=0$ has a polynomial solution of degree $n\geq0$ if and only if
$\mathit{rank(A_{n+m})=rank(A_{n+m-1})}$\textit{$.$} In this case, since
$A_{n+m}$ is upper triangular, the last entry of $A_{n+m}$ is zero i.e.
$\displaystyle\lambda_{n+m}=\sum_{k\geq1}a_{kk}(m+n-k)_{k}=0,$ and therefore
the last row of $A_{n+m}$ is zero.\newline

\noindent Now let $M_{n}$ and $M_{n}^{\prime}$ be, respectively, the matrices
obtained from $A_{n+m}$ and $A_{n+m-1}$ by deleting the first $m$ zero
columns. Clearly $rank(A_{n+m})=rank(A_{n+m-1})$ if and only if $rank(M_{n}%
)=rank(M_{n}^{\prime})$. It is easy to see that the $(i,j)^{th}$ entry of the
$(n+m+1)\times(n+1)$ matrix $M_{n}$ is $\displaystyle\sum_{t=0}^{j-1}%
a_{t+m,\,t+i-j}(j-t)_{t+m}=\sum_{t=0}^{j-1}p_{t,\,t+i-j}(j-t)_{t+m}.$ This
proves the main result of this note:\newline

\noindent {\bf Proposition 2.} Let $M_{n}$ be the $(n+m+1)\times(n+1)$ matrix with
$(i,j)^{th}$ entry $\displaystyle\sum_{t=0}^{j-1}p_{t,\,t+i-j}(j-t)_{t+m}$ and
let $M_{n}^{\prime}$ be the matrix obtained from $M_{n}$ by deleting the last
column. Then the differential equation $Ly=0$ has a polynomial solution of
degree $n\geq0$ if and only if $rank(M_{n})=rank(M_{n}^{\prime})$.$\Box
$\newline

\noindent It thus follows that if the equation $Ly=0$ has a polynomial
solution of degree $n\geq0$, then $\lambda_{n+m}=\displaystyle\sum_{t\geq
1}p_{t,\,t+m}(n-t)_{({t+m})}=0$, and since $M_{n}^{\prime}$ has $n$ columns,
$rank(M_{n})=rank(M_{n}^{\prime})$ implies that $rank(M_{n})\leq n$ and so
every $(n+1)\times(n+1)$ submatrix of $M_{n}$ has zero determinant. This
generalizes Theorems 5 and 6 of \cite{ciftci2010}.\newline

\section{Maple Code and Examples}

\noindent
In this section we illustrate the effectiveness of the algorithmic approach of
Section 2 by applying it to four differential equations that appear in
\cite{ciftci2010}. These arise in mathematical physics and, more precisely, in
the study of solutions to Schr\"{o}dinger equation \cite{C11}, planar Coulomb
diamagnetic problem \cite{C10}, Bohr Hamiltonian with Davidson potential
\cite{C15} and radial Schr\"{o}dinger equation with shifted potential
\cite{ciftci2010, C27, C28}. The examples show how to implement the method
algorithmically to determine the conditions for the existence of polynomial
solutions and also to calculate the corresponding polynomial solutions.
\newline

\noindent\textbf{Example 1}\newline As a first example, we consider the linear
second order ODE arising in the study of one dimensional Schr\"{o}dinger
problems \cite{C11}. The investigation of Krylov and Robnik \cite{C11} about
polynomial solutions of one dimensional Schr\"{o}dinger problems leads to
investigation of polynomial solutions of the following differential equation.
The conditions for existence of polynomial solutions of this ODE have also
been discussed by Ciftci et al. in the recent paper \cite{ciftci2010}, by a
different approach.
\begin{equation}
{x}^{3}{\frac{d^{2}}{d{x}^{2}}}y\left(  x\right)  +\alpha\,\left(  {x}%
^{2}-1\right)  {\frac{d}{dx}}y\left(  x\right)  +\left(  \beta\,x+g\right)
y\left(  x\right)  =0 \label{ODE-eg1}%
\end{equation}
Here, we apply our method to determine existence conditions as well as to
compute the corresponding polynomial solutions of the above differential
equation. A Maple code is also provided to show how the method can be
implemented algorithmically using software.

For $g=0$, the algorithmic procedure of Section 2 can be implemented to
generate a sequence of even degree polynomial solutions of ODE~(\ref{ODE-eg1}%
). A further analysis of these solutions yields the following result.

\begin{itemize}
\item For $g= 0$, ODE~(\ref{ODE-eg1}) admits polynomial solutions of degree
$n=2m$ $(m\geq1)$, with $\beta= -(\alpha n + n^{2} -n)$, given by
\[
y=x^{2m}+\sum_{i=1}^{m}\frac{(-1)^{i} {\binom{m }{{i}}} \alpha^{i} x^{2m-2i}%
}{(\alpha+2n-3)(\alpha+2n-5)\cdots(\alpha+2n-3-2(i-1))}
\]
It should be noted that, for $g=0$, no lower odd degree polynomial solutions
of ODE~(\ref{ODE-eg1}) were found.
\end{itemize}

For $g\neq0$ some examples of polynomial solutions, found using the
construction procedure of Section 2, are given in Table 1 below.

\begin{table}[h]
\begin{tabular}
[c]{|l|l|l|l|}\hline
$n$ & $\beta$ & $g$ & Polynomial solution of ODE~(\ref{ODE-eg1}) of degree
$n$\\\hline
$1$ & $-\alpha$ & $\pm\alpha$ & $x\pm1$\\\hline
&  &  & \\
$2$ & $-2\alpha-2$ & $\pm\sqrt{4\alpha^{2} +6\alpha}$ & $x^{2} \pm\frac
{\sqrt{2\alpha(2\alpha+3)}}{\alpha+2} + \frac{\alpha}{\alpha+2}$\\\hline
&  &  & \\
$3$ & $-3\alpha-6$ & $\pm\sqrt{15\alpha+5\alpha^{2} +\alpha A} $ & $4x^{3}
\pm\frac{12\alpha(9+2\alpha+A)}{\alpha\sqrt{15+5\alpha+A} (-3-2\alpha+A)}%
x^{2}$\\
&  &  & $+\frac{24\alpha}{-3-2\alpha+A}x \pm\frac{24\alpha^{2}}{\alpha
(-3-2\alpha+A)\sqrt{15+5\alpha+A}} $\\\cline{3-4}
&  &  & \\
&  & $\pm\sqrt{15\alpha+5\alpha^{2} -\alpha A} $ & $4x^{3} \pm\frac
{12\alpha(-9-2\alpha+A)}{\alpha\sqrt{15+5\alpha-A} (3+2\alpha+A)}x^{2}$\\
&  &  & $-\frac{24\alpha}{3+2\alpha+A}x \mp\frac{24\alpha^{2}}{\alpha
(3+2\alpha+A)\sqrt{15+5\alpha-A}} $\\
&  &  & where\\
&  &  & $A=\sqrt{153+96\alpha+16\alpha^{2}}$\\\hline
\end{tabular}
\caption{}
\end{table}

In general, for any given $n$ and $\alpha$, the algorithmic procedure of
Section 2 can easily be implemented to determine $\beta$, $g$ for which
ODE~(\ref{ODE-eg1}) admits polynomial solutions of degree $n$ as well as to
compute the corresponding polynomial solution. We provide below a set of Maple
commands that can be used as template to compute polynomial solutions of given
degree $n$ of ODE~(\ref{ODE-eg1}) for any given value of $\alpha$. For
illustration we take $\alpha=\frac{-15}{2}$ and look for a solution of degree
$n=6$. The following Maple code determines $\beta=15$, $g=3(750)^{1/4}$ and
computes the corresponding polynomial solution of degree $6$ of
ODE~(\ref{ODE-eg1}) as
\begin{align}
y(x)  &  =7\,{x}^{6}+{\frac{7}{30}}\,{1080}^{3/4}{x}^{5}+126\,{\frac{\sqrt
{30}\left(  -65+12\,\sqrt{30}\right)  }{66\,\sqrt{30}-360}}{x}^{4}%
+420\,{\frac{\sqrt[4]{1080}\left(  6\,\sqrt{30}-30\right)  }{66\,\sqrt
{30}-360}}{x}^{3}\nonumber\label{sol-1-n-6}\\
&  +1575\,{\frac{\left(  -72+6\,\sqrt{30}\right)  }{66\,\sqrt{30}-360}}{x}%
^{2}+630\,{\frac{\left(  -72+6\,\sqrt{30}\right)  {750}^{3/4}}{\left(
66\,\sqrt{30}-360\right)  \left(  -60+5\,\sqrt{30}\right)  }}x\\
&  -7875\,{\frac{\left(  -72+6\,\sqrt{30}\right)  \sqrt{30}}{\left(
66\,\sqrt{30}-360\right)  \left(  -60+5\,\sqrt{30}\right)  }}\nonumber
\end{align}
The Maple code with brief explanations is presented below. \newline%
\textit{\indent restart:\newline\indent with(LinearAlgebra):\newline%
\indent$\alpha:=-15/2:$ \newline\indent$N:=2:$\newline\indent pcoeff :=
Array(0 .. N):\newline\indent pcoeff[0] := $\beta x+g:$ \newline\indent
pcoeff[1] := $\alpha(x^{2}-1):$\newline\indent pcoeff[2] := $x^{3}:$\newline}
These commands define the order $N$ of the ODE, the value of $\alpha$ and the
coefficients $p_{k}(x)$ of the operator $\sum_{k=0}^{N}D^{k}y$. Next we
determine the value of $m$ and the coefficients $a_{k}(x)$ of the operator
$\sum_{k=1}^{m+N}D^{k}z$ using the following commands.\newline%
\textit{\indent$Vm:=Array(0..N):$\newline\indent$dk:=Array(0..N):$
\newline\indent for i from 0 to N do \newline\indent\quad dk[i] :=
degree(pcoeff[i], x) \newline\indent\quad Vm[i]:= degree(pcoeff[i],x)-i end
do:\newline\indent$m:=max(Vm):$\newline\indent$dkmax:=max(dk):$ \newline%
\indent acoeff := Array(1 .. m+N):\newline\indent for i from m to (m+N)
do\newline\indent\quad acoeff[i]:= pcoeff[i-m] end do:\newline} Matrices
$A_{n+m}$, $A_{n+m-1}$ of the procedure of Section 2 are computed by the
following set of commands in which $"soldegree"$, $An$ and $Anprime$
respectively denote the degree of the sought polynomial solution, the matrix
$A_{n+m}$ and the matrix $A_{n+m-1}$.\newline\textit{\indent soldegree :=
6:\newline\indent n := soldegree+m:\newline\indent cAM := max(dkmax,
n+m+N):\newline\indent rAM := m+N:\newline\indent AM := Array(1 .. rAM, 0 ..
cAM):\newline\indent for i from m to rAM do\newline\indent\quad AM[i,0]:=
coeff(x acoeff[i],$x^{1}$) end do:\newline\indent for i from m to rAM do
\newline\indent\quad for j from 1 to cAM do \newline\indent\qquad AM[i, j] :=
coeff(acoeff[i], $x^{j}$)\newline\indent\quad end do \newline\indent end
do\newline} \textit{\indent An := Matrix(n+1,n+1):\newline\indent for i from 1
to n+1 do \newline\indent\quad for j from 1 to n+1 do\newline\indent\qquad for
k from 1 to rAM do \newline\indent\qquad\quad if (k+i-j) $\geq0$ and (k-j+1)
$\leq0$ \newline\indent\qquad\qquad then An[i,j]:= An[i,j]+AM[k,k+i-j]
$\frac{(j-1)!}{(j-k-1)!}$ \newline\indent\qquad\quad end if \newline%
\indent\qquad end do\newline\indent\quad end do \newline\indent end do
\newline} \textit{\indent Anprime := Matrix(n+1,n) \newline\indent for i from
1 to n+1 do \newline\indent\quad for j from 1 to n do\newline\indent\qquad
Anprime[i, j] := An[i, j] \newline\indent\quad end do \newline\indent end do
\newline} At this stage we have $Rank(An)\neq Rank(Anprime)$. Next we
determine the values of parameters so that $Rank(An)$ equals $Rank(Anprime)$.
The first condition employed is the vanishing of the last diagonal entry of
the upper triangular matrix $An$ which determines $\beta$ by the following
commands. \textit{\indent ansbeta := solve(An[n+1, n+1] = 0, $\beta$):
\newline\indent$\beta$ := ansbeta \newline} Implementing the fact that the
$7\times7$ submatrix, obtained by deleting the last zero row of $An$, must
have zero determinant provides the value of $g$ via the commands
below.\newline\textit{\indent Andet := Matrix(n,n):\newline\indent for i from
1 to n do \newline\indent\quad for j from 1 to n do \newline\indent\qquad
Andet[i, j] := An[i, j+1] \newline\indent\quad end do \newline\indent end
do\newline\indent Determinant(Andet):\newline\indent ansdet :=
solve(Determinant(Andet) = 0, g)\newline} This leads to seven roots. At this
stage a root needs to be chosen before checking the rank condition. For
illustration we choose $g=3(750)^{1/4}.$\newline\textit{\indent g:=
$3(750)^{1/4}:$ \newline\indent Rank(An) \newline\indent Rank(Anprime)
\newline} As the rank condition is satisfied so the desired polynomial
solution can be obtained by the following set of commands.\newline%
\textit{\indent kern := NullSpace(An) \newline} The output of the above
command contains two vectors $kern_{1}$ and $kern_{2}$ with $kern_{2}%
=(1,0,0,0,0,0,0)^{T}$ so the vector $kern_{1}$ is used as follows to find the
solution.\newline\textit{\indent Vkern := $kern_{1}:$ \newline\indent solm :=
0: \newline\indent for i from 1 to (n+1) do \newline\indent\quad solm:= solm +
Vkern[i] $x^{i-1}$ end do: \newline\indent soln := diff(solm, x) \newline} The
output \textit{soln} provides the solution given in Equation~(\ref{sol-1-n-6}%
).\newline

\noindent\textbf{Example 2}\newline Consider the ODE
\begin{equation}
{\frac{d^{2}}{d{x}^{2}}}y\left(  x\right)  +\left(  p-2{x}^{2}\right)
{\frac{d}{dx}}y\left(  x\right)  +\left(  \delta\,x+\alpha\right)  y\left(
x\right)  =0 \label{ODE-eg2}%
\end{equation}
The question of investigating the polynomial solutions of ODE~(\ref{ODE-eg2})
arises from the study of polynomial solutions of Coulomb diamagnetic problem
by Chhajlany and Malnev \cite{C10}. Ciftci et al. \cite[Eq.18,19]{ciftci2010}
provide conditions for the existence of polynomial solutions of
ODE~(\ref{ODE-eg2}). Here we implement our procedure to demonstrate how to
generate polynomial solutions of ODE~(\ref{ODE-eg2}) in a straightforward
algorithmic manner.

For general $\alpha\neq0$ some examples of polynomial solutions listed in
Table 2 below are obtained by adapting the Maple code presented in Example 1.
The conditions on the parameters $\delta$ and $p$, for having these solutions,
are also determined.

\begin{table}[h]
\begin{tabular}
[c]{|l|l|l|l|}\hline
$n$ & $\delta$ & $p$ & Polynomial solution of ODE~(\ref{ODE-eg2}) of degree
$n$\\\hline
$1$ & $2$ & $\frac{\alpha^{2}}{2}$ & $2x-\alpha$\\\hline
&  &  & \\
$2$ & $4$ & $\frac{\alpha^{3} + 16}{8\alpha}$ & $x^{2}- \frac{\alpha}{2}x +
{\frac{{\alpha}^{3}-16}{16 \alpha}}$\\\hline
&  &  & \\
$3$ & $6$ & ${\frac{5\alpha^{2}}{18}}\,\pm\frac{2 \sqrt{{\alpha}%
^{4}-54\,\alpha}}{9}\,$ & $4x^{3} -2\alpha x^{2} - \frac{1}{3}{\frac{ \left(
\pm5\,{\alpha}^{3}+4\,\sqrt{\alpha\, \left(  {\alpha}^{3} -54 \right)  }%
\alpha\mp216 \right)  \alpha\,x}{{\pm\alpha}^{2}+2\,\sqrt{ \alpha\, \left(
{\alpha}^{3}-54 \right)  }}}$\\
&  &  & $+ \frac{1}{54} {\frac{\pm41\,{\alpha}^{5}+40\,{\alpha}^{3}%
\sqrt{\alpha\, \left(  { \alpha}^{3}-54 \right)  } \mp1728\,{\alpha}%
^{2}-432\,\sqrt{\alpha\, \left(  {\alpha}^{3}-54 \right)  }}{{\pm\alpha}%
^{2}+2\,\sqrt{\alpha\, \left(  {\alpha}^{3}-54 \right)  }}} $\\\hline
&  &  & \\
4 & 8 & ${\frac{1}{64}}\,{\frac{5\,{\alpha}^{3}+192\pm3\,A}{\alpha}} $ &
$5x^{4} - \frac{5\alpha}{2}x^{3} + {\frac{15}{32}}\,{\frac{ \left(  \mp
{\alpha}^{6}\pm768\,{\alpha}^{3}+{ \alpha}^{3}A\mp4096-64\,A \right)  {x}^{2}%
}{\alpha\, \left(  \pm3\,{\alpha}^{ 3} \mp192+5\,A \right)  }}$\\
&  &  & $\pm{\frac{5}{64}}\,{\frac{ \left(  \mp{\alpha}^{9}\mp128\,{\alpha
}^{6}+{ \alpha}^{6}A\pm8192\,{\alpha}^{3}-1728\,{\alpha}^{3}A\pm262144+4096\,A
\right)  x}{ \left(  \pm3\,{\alpha}^{3}\mp192+5\,A \right)  \left(  \mp
{\alpha} ^{3}\pm64+A \right)  }} $\\
&  &  & $\mp{\frac{5}{2048}}\,{\frac{B}{{\alpha}^{2} \left(  \pm3\,{\alpha
}^{3}\mp192 +5\,A \right)  \left(  \mp{\alpha}^{3}\pm64+A \right)  }} $\\
&  &  & where\\
&  &  & $A=\sqrt{{\alpha}^{6}- 384\,{\alpha}^{3}+4096}$ and\\
&  &  & $B = \mp{\alpha}^{12}\mp2816\,{\alpha}^{9}+{\alpha}^{9}A\pm
524288\,{\alpha}^{6}- 5184\,{\alpha}^{6}A $\\
&  &  & $\pm19922944\,{\alpha}^{3}+200704\,{\alpha}^{3}A\mp
150994944-2359296\,A$.\\\hline
\end{tabular}
\caption{}
\end{table}

\vspace{1.1 in}
Depending on the available computational power, for a given value of $\alpha$,
the algorithmic procedure of Section 2 can be implemented, as in Example 1, to
compute polynomial solutions of ODE~(\ref{ODE-eg2}) of any given degree $n$.
As example the following solutions of degree 9 and 25 are found.

\begin{itemize}
\item $\alpha=0$, $n=9$ implies $\delta= 18$, $p=\frac{4}{35}(144830)^{1/3}$
and the polynomial solution of degree $9$ of ODE~(\ref{ODE-eg2}) given by
\begin{align*}
y(x)  &  = 10 x^{9} -{\frac{18}{7}}\,\sqrt[3]{144830}{x}^{7} - 120 x^{6} +
{\frac{9}{35}}\,{144830}^{2/3}{x}^{5} + {\frac{666}{35}}\,\sqrt[3]{144830}%
{x}^{4}\\
&  - {\frac{10314}{7}}\,{x}^{3} -{\frac{8478}{8575}}\,{144830}^{2/3}{x}^{2} +
{\frac{4239}{245}}\,\sqrt[3]{144830}x + {\frac{19803534}{8575}}%
\end{align*}

\item For $\alpha=0$, $n=25$ with $\delta=50$ and $p=0$ the polynomial
solution of degree $25$ of ODE~(\ref{ODE-eg2}) is given by
\begin{align*}
y(x)  &  =26\,{x}^{25}-2600\,{x}^{22}+100100\,{x}^{19}-1901900\,{x}%
^{16}+19019000\,{x}^{13}\\
&  -98898800\,{x}^{10}+247247000\,{x}^{7}-247247000\,{x}^{4}+61811750\,x
\end{align*}
\end{itemize}

\noindent\textbf{Example 3}\\
The analysis of solutions of the Bohr Hamiltonian for Davidson potential leads
to investigation of exact solutions of a differential equation \cite[Eq.49]%
{C15} which can be rewritten as \cite[Eq.22]{ciftci2010}
\begin{equation}
x{\frac{d^{2}}{d{x}^{2}}}y\left(  x\right)  -\left(  2\,{x}^{2}-2\,\mu
-2\right)  {\frac{d}{dx}}y\left(  x\right)  -\left(  2\,\mu+3-\epsilon\right)
xy\left(  x\right)  =0 \label{ODE-eg3}%
\end{equation}
Adapting the Maple code of Example 1 for ODE~(\ref{ODE-eg3}) readily generates
polynomial solutions of a given degree $n$. Some examples for solutions of
even as well as odd degrees are provided below in Tables 3 and 4
respectively.\newline

\begin{table}[h]%
\begin{tabular}
[c]{|l|l|l|}\hline
$n$ & $\epsilon$ & Polynomial solution of ODE~(\ref{ODE-eg3}) of degree
$n=2m$\\\hline
$0$ & $2\mu+3$ & $1$\\\hline
$2$ & $2\mu+7$ & $2\,{x}^{2}-\,(2\mu+3)$\\\hline
$4$ & $2\mu+11$ & $4\,{x}^{4} - 4\, \left(  2\,\mu+5 \right)  {x}^{2}+ \left(
2\,\mu+3 \right)  \left(  2\,\mu+5 \right)  $\\\hline
$6$ & $2\mu+15$ & $8\,{x}^{6}- 12\, \left(  2\,\mu+7 \right)  {x}^{4}$\\
&  & $+6\, \left(  2\,\mu+5 \right)  \left(  2\,\mu+7 \right)  {x}^{2}-
\left(  2\,\mu+3 \right)  \left(  2\,\mu+5 \right)  \left(  2\,\mu+7 \right)
$\\\hline
$8$ & $2\mu+19$ & $16\,{x}^{8}-32\, \left(  2\,\mu+9 \right)  {x}^{6}+24\,
\left(  2\,\mu+7 \right)  \left(  2\,\mu+9 \right)  {x}^{4}$\\
&  & $-8\, \left(  2\,\mu+5 \right)  \left(  2\,\mu+7 \right)  \left(
2\,\mu+9 \right)  {x}^{2}+ \left(  2\,\mu+3 \right)  \left(  2\, \mu+5
\right)  \left(  2\,\mu+7 \right)  \left(  2\,\mu+9 \right)  $\\\hline
$10$ & $2\mu+23$ & $32\,{x}^{10}-80 \, \left(  2\,\mu+ 11 \right)  {x}%
^{8}+80\, \left(  2\,\mu+9 \right)  \left(  2\,\mu+ 11 \right)  {x}^{6}$\\
&  & $-40\, \left(  2\,\mu+7 \right)  \left(  2\,\mu+9 \right)  \left(
2\,\mu+11 \right)  {x}^{4} $\\
&  & $+10\, \left(  2\,\mu+5 \right)  \left(  2\,\mu+7 \right)  \left(
2\,\mu+9 \right)  \left(  2\, \mu+11 \right)  {x}^{2}$\\
&  & $- \left(  2\,\mu+3 \right)  \left(  2\,\mu+5 \right)  \left(  2\,\mu+7
\right)  \left(  2\,\mu+9 \right)  \left(  2\, \mu+11 \right)  $\\\hline
\end{tabular}
\caption{}
\end{table}

\noindent It should be noted that for this case polynomial solutions can
easily be computed without much computational cost. For instance, Maple could
compute polynomial solution of degree $n=100$ in computational time of $1.8$
seconds. \newline

\begin{table}[h]%
\begin{tabular}
[c]{|l|l|l|l|}\hline
$n$ & $\epsilon$ & $\mu$ & Polynomial solution of ODE~(\ref{ODE-eg3}) of
degree $n=2m+1$\\\hline
$1$ & $2\mu+5$ & $-1$ & $x$\\\hline
$3$ & $2\mu+9$ & $-1$ & $x^{3} - \frac{3}{2}x $\\\cline{3-4}
&  & $-2$ & $x^{3}$\\\hline
$5$ & $2\mu+13$ & $-1$ & $x^{5} - 5 x^{3} + \frac{15}{4}x $\\\cline{3-4}
&  & $-2$ & $x^{5}- \frac{5}{2}x^{3}$\\\cline{3-4}
&  & $-3$ & $x^{5}$\\\hline
$7$ & $2\mu+17$ & $-1$ & $x^{7} - \frac{21}{2} x^{5} + \frac{105}{4}x^{3} -
\frac{105}{8} x $\\\cline{3-4}
&  & $-2$ & $x^{7} - 7 x^{5} + \frac{35}{4}x^{3}$\\\cline{3-4}
&  & $-3$ & $x^{7} - \frac{7}{2} x^{5}$\\\cline{3-4}
&  & $-4$ & $x^{7}$\\\hline
\end{tabular}
\caption{}
\end{table}

\noindent A further analysis of the odd lower degree polynomial solutions of
ODE~(\ref{ODE-eg3}), given in Table 4, yields the following family of
solutions of degree $n=2m+1$. \newline For $k=0,1,2,3$ ODE~(\ref{ODE-eg3})
admits the following class of polynomial solutions of degree $n=2m+1$ $(m\geq
k)$ with $\mu=-(m+1-k)$ and $\epsilon=2m+2k+3$.

\begin{itemize}
\item If $k=0$ $(m\geq0)$
\[
y=x^{2m+1}%
\]

\item if $k=1$ $(m\geq1)$
\[
y=x^{2m+1}-\frac{2m+1}{2}x^{2m-1}%
\]

\item if $k=2,3$ $(m\geq k)$
\begin{align*}
y  &  = x^{2m+1} + k \sum_{i=1}^{k-1}\frac{(-1)^{i} (2m+1)(2m-1)\cdots
(2m+1-2(i-1))}{2^{i}}x^{2m+1-2i}\\
&  + \frac{(-1)^{k} (2m+1)(2m-1)\cdots(2m+1-2(k-1))}{2^{k}}x^{2m+1-2k}%
\end{align*}

\end{itemize}

\noindent\textbf{Example 4}\newline As a final example, we consider a question
related to the investigation of the radial Schr\"{o}dinger equation with
shifted Coulomb potential which has been discussed recently in
\cite{ciftci2010, C27, C28}. The anstaz of \cite[Eq.35]{ciftci2010} that the
radial Schr\"{o}dinger equation admits a solution which vanishes at the origin
and at infinity leads to the question of obtaining solutions of the following
differential equation; the reader is referred to \cite{ciftci2010} for
details.
\begin{align}
&  x(x+\beta){\frac{d^{2}}{d{x}^{2}}}y(x)+\left(  -2\alpha{x}^{2}%
+2(K+1-\alpha\beta)x+2\beta(K+1)\right)  {\frac{d}{dx}}%
y(x)\nonumber\label{ODE-eg4}\\
&  +\left(  (-2\alpha(K+1)+2Z)x-2\alpha\beta(K+1)\right)  y(x)=0
\end{align}
This is a particular case of the confluent Heun equation whose polynomial
solutions can be studied algorithmically using our procedure. While discussing
the question of polynomial solutions of ODE~(\ref{ODE-eg4}), Ciftci et al.
\cite{ciftci2010} provide conditions on parameters $\alpha,\ \beta$ to have
polynomial solutions. In particular a table was provided which listed
conditions on parameters for the existence of polynomial solutions for
$n=1,2,3,4.$ However, it was pointed out in \cite{ciftci2010} that finding the
corresponding polynomial solutions is an open problem that remains to be
solved. For a given value $K$ and the given degree $n$ of the required
polynomial solution, adapting the Maple code of example 1 for
ODE~(\ref{ODE-eg4}) can determine conditions on parameters $\alpha,\ \beta$ as
well as generate corresponding polynomial solutions of ODE~(\ref{ODE-eg4}). In
Table 5 below, we demonstrate this by providing some examples of polynomial
solutions of ODE~(\ref{ODE-eg4}) of degree $n=1,2,3,4,5$.

\begin{table}[h]%
\begin{tabular}
[c]{|l|l|l|l|l|}\hline
$n$ & $\alpha$ & $K$ & $\beta$ & Polynomial solution of ODE~(\ref{ODE-eg4}) of
degree $n$\\\hline
$1$ & $\frac{Z}{K+2}$ & any & $\frac{K+2}{Z}$ & $x+ \frac{K+2}{Z}$\\\hline
$2$ & $\frac{Z}{K+3}$ & any & $\,{\frac{ \left(  3\,K+6\pm\sqrt{{K}%
^{2}+8\,K+12} \right)  \left(  K+ 3 \right)  }{ 2\left(  K+2 \right)  Z}} $ &
$3\,{x}^{2}+3\,{\frac{ \left(  K+3 \right)  \left(  3\,K+6\pm\sqrt{{K}^{2
}+8\,K+12} \right)  \left(  2\,K+3 \right)  x}{ \left(  K+2 \right)  Z \left(
K\pm\sqrt{{K}^{2}+8\,K+12} \right)  }} $\\
&  &  &  & $+3\,{\frac{ \left(  3\,K+6 \pm\sqrt{{K}^{2}+8\,K+12} \right)
\left(  K+3 \right)  ^{2} \left(  2\,K+ 3 \right)  }{ \left(  K+2 \right)
{Z}^{2} \left(  K\pm\sqrt{{K}^{2}+8\,K+ 12} \right)  }} $\\\hline
$3$ & $\frac{Z}{K+4}$ & $-\frac{3}{2}$ & $\frac{5}{2Z}$ & $4x^{3}+ \frac
{10}{Z}x^{2}$\\\cline{4-5}
&  &  & $\frac{25}{2Z}$ & $4x^{3}+ \frac{110}{Z}x^{2} + \frac{1875}{2Z^{2}}x +
\frac{9375}{4Z^{3}}$\\\hline
$4$ & $\frac{Z}{K+5}$ & $-\frac{3}{2}$ & $\frac{49}{2Z}$ & $5{x}^{4}%
+{\frac{1435}{4}}\,{\frac{{x}^{3}}{Z}}+{\frac{18375}{2}} \,{\frac{{x}^{2}}%
{{Z}^{2}}}+{\frac{3109295}{32}}\,{\frac{x}{{Z}^{3} }}+{\frac{21765065}%
{64Z^{4}}}$\\\cline{4-5}
&  &  & $\frac{7(\pm15+\sqrt{65})}{20Z}$ & $5{x}^{4}+28\,{\frac{ \left(
\pm15+\sqrt{65} \right)  {x}^{3}}{Z \left(  \pm1+\sqrt{65} \right)  }%
}+98\,{\frac{ \left(  \pm15+\sqrt{65} \right)  {x}^{2}}{ \left(  \pm
1+\sqrt{65} \right)  {Z}^{2}}} $\\\hline
$5$ & $\frac{Z}{K+6}$ & $-\frac{3}{2}$ & $\frac{81}{2Z}$ & $6{x}%
^{5}+891\,{\frac{{x}^{4}}{Z}}+51030\,{\frac{{x}^{3}}{{Z}^{2}}}
+1390932\,{\frac{{x}^{2}}{{Z}^{3}}}$\\
&  &  &  & $+{\frac{282195171}{16}}\,{\frac{ x}{{Z}^{4}}}+{\frac
{2539756539}{32Z^{5}}} $\\\hline
\end{tabular}
\caption{}
\end{table}

\end{document}